\documentclass[pra,showpacs,preprintnumbers,amsmath,amssymb,floatfix]{revtex4}
\usepackage{hyperref}
\usepackage[dvips]{epsfig}
\newcommand{\beq}{\begin{equation}}
\newcommand{\eeq}{\end{equation}} 
\newcommand{\beqa}{\begin{eqnarray}}
\newcommand{\eeqa}{\end{eqnarray}}
\newcommand{\ba}{\begin{array}}
\newcommand{\ea}{\end{array}}

\begin{document}

\title{Density functional theory of a trapped Bose gas with 
tunable scattering length: from weak-coupling to unitarity}
\author{Maurizio Rossi$^1$, 
         Francesco Ancilotto$^{1,2}$, 
         Luca Salasnich$^{1,3}$, 
         Flavio Toigo$^{1,2}$}
\affiliation{Dipartimento di Fisica e Astronomia "Galileo Galilei" 
and CNISM, Universit\`a di Padova, via Marzolo 8, 35122 Padova, Italy 
\\
CNR-IOM Democritos, via Bonomea, 265 - 34136 Trieste, Italy
\\
CNR-INO, via Nello Carrara, 1 - 50019 Sesto Fiorentino, Italy}

\begin{abstract}We study an interacting Bose gas at T=0 under isotropic 
harmonic confinement within Density Functional Theory in the Local 
Density approximation.
The energy density functional, which spans the whole range of positive 
scattering lengths up to the unitary regime (infinite scattering length), 
is obtained by fitting the recently calculated Monte Carlo bulk equation 
of state [Phys. Rev. A {\bf 89}, 041602(R) (2014)]. 
We compare the density profiles of the trapped gas with those obtained 
by MC calculations.
We solve the time-dependent Density Functional equation to study the 
effect of increasing values of the interaction strength on the 
frequencies of monopole and quadrupole oscillations of the trapped gas.  
We find that the monopole breathing mode shows a non-monotonous behavior
as a function of the scattering length. 
We also consider the damping effect of three-body losses on such modes.
\end{abstract}

\maketitle

Motivated by promising experimental observations \cite{liho,brem,corn,zora}, 
in a very recent paper \cite{noi1} we have investigated the zero-temperature 
properties of a dilute homogeneous Bose gas by tuning the interaction 
strength of the two-body potential to achieve arbitrary positive values of 
the s-wave scattering length $a$. 
In that paper \cite{noi1} we have computed by Monte Carlo (MC) quadrature 
the energy per particle and the condensate fraction of the system by using 
a Jastrow ansatz for the many-body wave function which avoids the formation 
of the self-bound clusters present in the ground-state and describes instead 
a (metastable) gaseous state with uniform density.
 
Here we set up a reliable energy density functional (DF) for bosons under 
external confinement by fitting the MC equation of state of the bulk 
system \cite{noi1}. 

The Kohn-Sham formulation\cite{kohn} of time-dependent Density Functional 
Theory (TDDFT)\cite{gross} for an inhomogeneous system of interacting bosons 
at T=0 (with local number density $n({\bf r})$ and mass $M$) is described, 
within the Local Density Approximation, by the equation:
\beq 
 i\hbar {\partial \Psi({\bf r},t)\over \partial t} = 
 \left[ -{\hbar^2 \nabla^2 \over 2M} + U({\bf r}) + 
 {\partial (n\varepsilon _a)\over \partial n}
 \right ] \Psi({\bf r},t)   
 \label{nlse}
\eeq
where $|\Psi ({\bf r})|^2=n({\bf r})$ and  $\varepsilon _a(n({\bf r}))$ is the 
energy per atom of a {\it homogeneous} system with density equal to the local 
density. 
Here $U({\bf r})$ describes the external confinement, which we assume to be
an isotropic harmonic potential,$U({\bf r})=\frac{1}{2}M\omega_H^2(x^2+y^2+z^2)$.
The associated total energy functional is
\beq 
 \label{dft}
 E = \int d^3{\bf r} \left\{\frac{\hbar^2}{2M}|\nabla \Psi({\bf r})|^2 
                                          + n({\bf r})\varepsilon _a(n({\bf r}))
                    + n({\bf r}) U({\bf r}) \right\}\ .
\eeq

As previously discussed, the values of $\varepsilon _a(n)$ have been recently 
computed with a MC approach \cite{noi1} for a wide range of (positive) values 
of the scattering length $a$ characterizing the interparticle interaction.
In the weakly interacting regime ($x\equiv a/r_0\ll  1$, where 
$r_0=(3/(4\pi n))^{1/3}$ is the average distance between bosons) the MC results 
for $\varepsilon _a(n)$ are very close to $\varepsilon _{\rm LHY}(n)$, the 
universal Bogoliubov prediction \cite{bogo} as corrected by Lee, Huang and 
Yang (LHY) \cite{lhyc}.
In the strong-coupling regime ($x\gg 1$), instead, MC data reach a plateau and, 
in the unitary limit ($a\to \infty$), a finite and positive energy per particle 
is found, $E/N=0.70 \ \varepsilon_B(n)$, 
where $\varepsilon_B(n)=\frac{\hbar^2}{2M}(6\pi^2n)^{2/3}$.

The equation of state \cite{noi1} from such MC calculation can be well interpolated as:
\beq
 \label{fit}
 \frac{\varepsilon _a(n)}{\varepsilon_B(n)} = \left\{
  \begin{array}{lcc}
   f_{\rm LHY}(x) + a_3x^3                               & {\rm for} & x<0.3     \\

   c_7x^7+c_6x^6+c_5x^5+c_4x^4+c_3x^3+c_2x^2+c_1x+c_0    & {\rm for} & 0.3<x<0.5 \\
   b_0 + b_1\tanh\left(b_2/x - 1 \right)                 & {\rm for} & x>0.5
  \end{array}
 \right.
\eeq
with $a_3 = 0.21$, $b_0=0.45$, $b_1= -0.33$, $b_2=0.54$, 
$c_0 = 4.75$ , $c_1 = -99.72$, $c_2 = 890.68$, $c_3 = -4309.56$, $c_4 = 12268.41$,
$c_5 = -20488.00$, $c_6 = 18568.27$ and $c_7 = -7052.20$ \cite{nota}.
In (\ref{fit}), $f_{\rm LHY}(x) = \left(\frac{4}{3\pi^2}\right)^{1/3}
x[1+\frac{128}{15\sqrt{\pi}}\sqrt{\frac{3}{4\pi}}x^{3/2}]$ is the LHY 
correction to the Bogoliubov prediction.

Notice that in the deep weak-coupling regime Eq. (\ref{nlse}) reduces to the 
familiar Gross-Pitaevskii equation (GPE) \cite{gpe} since 
$\varepsilon _a(n) = {\cal E}_{\rm GPE}(n,a) \equiv 2\pi\hbar^2 an^2/M$. 

\begin{figure}
 \begin{center}
  \epsfig{file=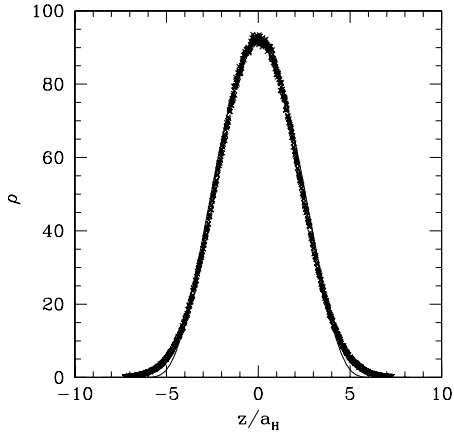,width=8.cm,clip=}    
  \vskip -0.8cm 
  \caption{Integrated density profile $\rho(z)$ for $N=500$ bosons with 
           $a=10^4\,a_0$ under isotropic harmonic confinement. 
           Solid line: DFT results; points: MC results. 
           Here $a_H\equiv \sqrt{\hbar /(M\omega_H)}$.}
  \label{fig1}
 \end{center}
 \vskip -0.4cm 
\end{figure}

In Fig.\ref{fig1} we show the ground-state integrated density profile 
$\rho(z)=\int dx dy \ n({x,y,z})$ for $N=500$ bosons and $a=10^4$ (in units 
of the Bohr radius $a_0$) obtained by numerically propagating Eq.(\ref{nlse}) 
in imaginary times. 
The DFT density profile is compared with the MC result for the same system. 
MC data are obtained by adding to the wave function of the bulk system, 
described in Ref.~\cite{noi1}, a standard Gaussian one-body term with a single 
variational parameter \cite{dubo}.
The comparison shows a quite good agreement apart near to the surface.
This discrepancy is mainly due to he fact that the value of the variational 
parameter in the one-body term, which is optimized by minimizing the energy 
per particle, is mainly determined by the higher density region at the center 
of the trap.  

\begin{figure}
 \begin{center} 
  \epsfig{file=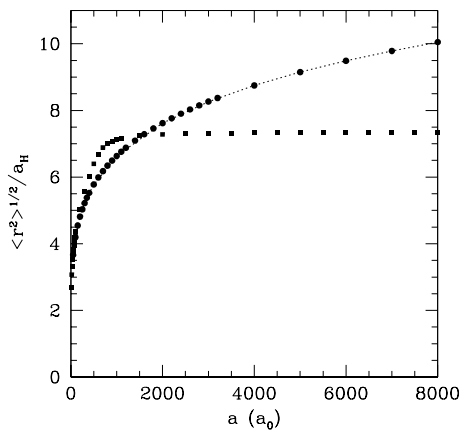,width=8.cm,clip=}
  \vskip -0.8cm 
  \caption{Average radius of the trapped gas as a function of $a$. 
           Squares: DFT results; dots: GPE results with the same value of $a$. 
           The dotted line shows the GPE result in the TF limit, where 
           $<r^2>^{1/2}\propto a^{1/5}$.}
  \label{fig2}
 \end{center}
 \vskip -0.4cm
\end{figure}

In Fig. \ref{fig2} we plot the average radius of the trapped gas as a function 
of the scattering length $a$. 
The figure clearly shows that DFT results (squares) converge to a finite 
radius for $a\gg 1$, while GPE one (dots and dotted line), obtained with 
$\varepsilon _a(n) = {\cal E}_{\rm GPE}(n,a)$, diverge as $a^{1/5}$
(Thomas-Fermi limit).
The convergence to a constant value is expected for the unitary regime, where 
the properties of the system become universal, i.e. depend only on the 
density and are insensitive to the actual value of $a$.

\begin{figure}
 \begin{center}
  \epsfig{file=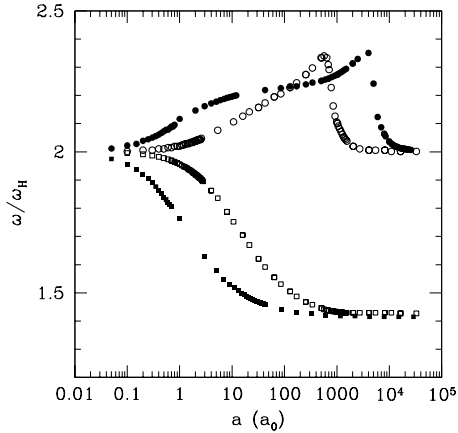,width=8.cm,clip=}
  \vskip -0.8cm 
  \caption{Monopole (upper curves) and quadrupole (lower curves) frequencies 
           $\omega$ as a function of $a$ obtained with DFT. 
           Empty symbols: $N=500$ bosons; filled symbols: $N=80000$ bosons.}
  \label{fig3}  
 \end{center}
 \vskip -0.4cm
\end{figure}

The oscillation frequencies of both monopole (breathing or compressional) and 
quadrupole (surface) modes for interacting bosons in harmonic trap can be 
obtained by numerically solving Eq. (\ref{nlse}).
The monopole mode is excited by slightly changing the frequency $\omega_H$ of 
the harmonic confinement when computing the initial state $\Psi({\bf r},t = 0)$, 
while the quadrupole mode is obtained by using as the initial state 
$\Psi({\bf r},t = 0) = e^{i\eta Q} \Psi_0({\bf r})$,  where $\Psi_0({\bf r})$ 
is the ground-state wave function, $\eta$ is a small parameter and 
$Q=2z^2-x^2-y^2$ is the standard quadrupole operator. 
The numerical results are shown in Fig. \ref{fig3} as a function of $a$.
As the scattering length $a$ is varied, the quadrupole frequency connects 
smoothly the values \cite{book} $2\omega_H$ appropriate to the non-interacting 
gas ($a\to 0$) and $\sqrt{2}\omega_H$ expected in the Thomas-Fermi (TF) regime 
($Na/a_H\gg 1$), where $a_H=\sqrt{\hbar/(M\omega_H)}$.
As expected, its TF value turns out to be independent of $a$ \cite{book}. 
The frequency of the monopole mode shows instead a non-monotonous behavior as a 
function of the scattering length. 
At small $a$ values (non interacting regime) it recovers the expected 
$2\omega_H$ limit \cite{book} and in the opposite limit ($a\to\infty$) it 
converges to the universal expected unitary value $2\omega_H$ \cite{cast}.
For intermediate values of $a$, the breathing frequency should approach the 
TF value $\sqrt{5}\omega_H$ (independent of $a$) \cite{book}; but this value is 
reached only in the $N=80000$ case, since for $N=500$ bosons the TF condition
$Na/a_H \gg 1$ is satisfied only for $a$ in the unitary regime (as inferred 
also from the slower convergence to the TF value of the quadrupole frequency).

\begin{figure}
 \begin{center}   
  \epsfig{file=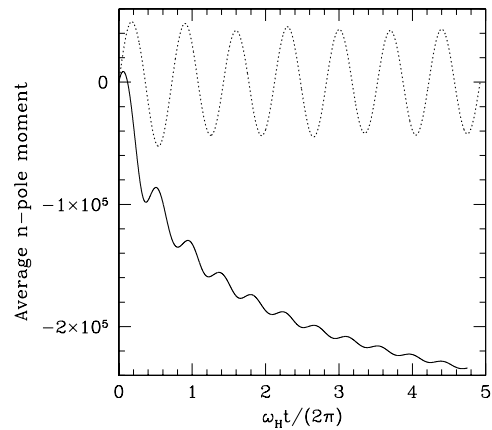,width=8.cm,clip=}
   \vskip -0.8cm
   \caption{Time evolution of the average $n-$pole moment of the trapped Bose 
            gas in the presence of the three-body loss term induced by monopole 
            (solid line) and quadrupole (dotted line) distorsions in the $N=80000$
            and $a=10^4\,a_0$ case.}
   \label{fig4}
  \end{center}
 \vskip -0.4cm
\end{figure}

Finally, we study the effect of three-body losses \cite{metz} on the collective 
oscillations by adding a dissipative term $-i\hbar L_3 n({\bf r},t)^2\Psi ({\bf r},t)$ 
(with $L_3=1.1\cdot 10^{-22}$ cm$^6$/sec \cite{corn}) in equation (\ref{nlse}). 
In Fig. \ref{fig4} we report the time evolution of the monopole and quadrupole mean
values, $<r^2>$ and $<Q>$ respectively, for $N=80000$ and $a=10^4\,a_0$. 
During the dynamics the number of atoms (and thus the volume) of the bosonic cloud
decreases and, as shown in the figure, $<r^2>$ follows the evolution of the average 
radius of the ground state with a superimposed oscillation with frequency close to 
$2\omega_H$.
On the contrary the surface mode is only slightly affected by the three-body losses, 
and $<Q>$ oscillates around its starting value with the frequency expected from the 
chosen value of $a$.

In conclusion, we have theoretically investigated statics and dynamics of a confined
Bose gas by means of a density functional approach which takes into account the whole 
range of positive scattering lengths from weak-coupling to unitarity. 
We have shown that the DFT density profiles of the ground state are in good agrement 
with MC ones. 
Moreover, by numerically solving the time-dependent density functional equation, 
we have calculated the frequencies of both monopole and quadrupole modes, 
analyzing also the effect of three-body losses.


\begin{thebibliography}{99}
\bibitem{nota} The coefficients $c$'s have been determined by ensuring continuity of
               the interpolating function and of its first three derivatives at $x=0.3$
               and $x=0.5$ once $a_3$ and the coefficients $b$'s have been fitted to the
               calculated values.
               This procedures automatically reproduces the MC values in the interval $0.3<x<0.5$.
\end{thebibliography}
\end{document}